\documentstyle[12pt]{article}
\makeatletter
\newcommand{\singlespacing}{\let\CS=\@currsize\renewcommand{\baselinestretch}
{1.0}\tiny\CS}
\newcommand{\doublespacing}{\let\CS=\@currsize\renewcommand{\baselinestretch}
{1.5}\tiny\CS}

\newcommand{\bd}{\begin{document}}
\newcommand{\ed}{\end{document}}
\newcommand{\bc}{\begin{center}}
\newcommand{\ec}{\end{center}}
\newcommand{\bfr}{\begin{flushright}}
\newcommand{\efr}{\end{flushright}}
\newcommand{\vs}{\vspace}
\newcommand{\hs}{\hspace}
\newcommand{\beq}{\begin{equation}}
\newcommand{\eeq}{\end{equation}}
\newcommand{\lb}{\linebreak}
\newcommand{\mb}{\makebox}
\newcommand{\fb}{\framebox}
\newcommand{\mc}{\multicolumn}
\newcommand{\ben}{\begin{enumerate}}
\newcommand{\een}{\end{enumerate}}
\newcommand{\bit}{\begin{itemize}}
\newcommand{\eit}{\end{itemize}}
\newcommand{\ol}{\overline}
\newcommand{\un}{\underline}
\newcommand{\lefq}{\lefteqn}
\newcommand{\ba}{\begin{array}}
\newcommand{\ea}{\end{array}}
\newcommand{\beqa}{\begin{eqnarray}}
\newcommand{\eeqa}{\end{eqnarray}}
\newcommand{\beqas}{\begin{eqnarray*}}
\newcommand{\eeqas}{\end{eqnarray*}}
\newcommand{\bfg}{\begin{figure}}
\newcommand{\efg}{\end{figure}}
\newcommand{\bds}{\begin{displaymath}}
\newcommand{\eds}{\end{displaymath}}
\newcommand{\btb}{\begin{tabbing}}
\newcommand{\etb}{\end{tabbing}}
\newcommand{\para}{\parallel}
\newcommand{\pad}{\partial}
\newcommand{\nn}{\nonumber}
\newcommand{\la}{\leftarrow}
\newcommand{\ra}{\rightarrow}
\newcommand{\lgla}{\longleftarrow}
\newcommand{\lgra}{\longrightarrow}
\newcommand{\La}{\Leftarrow}
\newcommand{\Ra}{\Rightarrow}
\newcommand{\Lra}{\Leftrightarrow}
\newcommand{\Lgla}{\Longleftarrow}
\newcommand{\Lgra}{\Longrightarrow}
\newcommand{\bm}{\boldmath}
\newcommand{\lan}{\langle}
\newcommand{\ran}{\rangle}
\renewcommand{\a}{\alpha}
\renewcommand{\b}{\beta}
\newcommand{\g}{\gamma}
\newcommand{\G}{\Gamma}
\renewcommand{\d}{\delta}
\newcommand{\eps}{\epsilon}
\newcommand{\th}{\theta}
\newcommand{\Th}{\Theta}
\newcommand{\s}{\sigma}
\newcommand{\lam}{\lambda}
\newcommand{\D}{\Delta}
\newcommand{\vare}{\varepsilon}
\newcommand{\pr}{\prime}
\newcommand{\ro}{\rho}
\newcommand{\nab}{\nabla}
\newcommand{\m}{\mu}
\newcommand{\n}{\nu}
\newcommand{\Sg}{\Sigma}
\newcommand{\p}{\pi}
\newcommand{\R}{I\!\!R}
\newcommand{\om}{\omega}
\newcommand{\Om}{\Omega}
\newcommand{\ze}{\zeta}
\newcommand{\vart}{\vartheta}
\newcommand{\tri}{\triangle}
\newcommand{\f}{\frac}
\newcommand{\iny}{\infty}
\newcommand{\pro}{\propto}
\begin{document}

\begin{center}{\large{\bf Topological Aspects of Spin Pairing in 
High-$T_c$ Superconductivity and Berry Phase}}\end{center}
\begin{center}
{\bf D.Pal}\
\footnote{e-mail : res9719@www.isical.ac.in}~and~{\bf B.Basu}\footnote
{ e-mail : banasri@www.isical.ac.in}\\
{\bf Physics and Applied Mathematics Unit}\\
{\bf Indian Statistical Institute}\\
{\bf Calcutta-700035}
\end{center}

\vspace{1 cm}

\centerline{\bf Abstract}
From topological viewpoint we have analysed  the role of Berry phase in 
spin pairing mechanism of high $T_c$ superconducting ground state.

\vspace{1 cm}

    Difficulty of achieving a high transition temperature for the high
temperature superconductors which are doped Mott
insulators via the conventional mechanism leads to a different approach 
towards
pairing. Any theory of high temperature superconductivity should satisfy the 
phenomenological constraint - the order parameter has charge 2e [1] i.e.
there is some kind of pairing. It is difficult to overcome the poor screening 
of  the Coulomb 
interaction of the doped Mott insulators  because of less effectiveness of 
retardation with the increment of the  pairing energy. In general the normal 
state has shown peculiar
properties in both charge and spin channels. In transport aspect, the linear
temperature dependence of resistivity in a long range of temperature 10K to
1000K, 1/T dependence and hole like charge carriers [2] of Hall coefficient,
$T^{-4}$ temperature dependence of magneto-resistance, monotonic decrease
and even sign changing of thermopower with increase of doping [3] are
anomalous with regard to the canonical phenomena in conventional Fermi liquid
(FL) system. Spin magnetic properties have also exhibited a number of
anomalies. Two powerful probes of spin dynamics in the cuprates - nuclear
magnetic resonance (NMR) [4] and neutron scattering [5] are inconsistent in
antiferromagnetic (AF) fluctuation case.
More generally, the conventional view of
superconductivity as a Fermi surface instability resulting from an attractive
interaction between quasiparticles is inapplicable, since according to
analyses of resistivity [6] and ARPES (angle resolved photoemission spectra)
data [7,8] there are no well defined quasiparticles or Fermi surface in the
normal state of high temperature superconductors. ARPES experiments which show
that the chemical potential is near the centre of the bare hole band, rule out
real-space pairing, which is implausible for d-wave superconductor with a
strong and poorly screened coulomb repulsion between electrons. In this note
we have investigated the relation of the antiferromagnetic insulating phase
and the superconducting phase in terms of Berry phase and hencce showed the
importance of this factor in spin pairing of the
superconducting state.

      For topological analysis of superconductivity let us  
 consider a 2D lattice system on the surface of a 3D sphere in an
anisotropic space with large radius. The antiferromagnetism may be
characterised by strong spin fluctuation in 2D which enables to consider the
system an anisotropic space. We may assume that this 2D lattice 
system is the CuO layer which plays a major role in the theory of high 
$T_c$ supereconductivity of cuprate materials. In a 3D anisotropic space, 
we can construct the spherical harmonics
$Y^{m,\m}_l$ with $l~=~1/2$, $|m|=|\m|=1/2$ when the angular momentum 
relation is given by [9]
\beq
\vec{J}~=~\vec{r}~\times \vec{p}~-~\m\vec{r}
\eeq
where $\m$ can take the values as $0,\pm 1/2,\pm 1,\pm 3/2, \ldots$
This is similar to the angular momentum relation in the case when a charged
particle moves in the field of a magnetic monopole. Fierz [10] and Hurst [11]
have studied the spherical harmonics $Y^{m,\m}_l$. Following them we can write
\beq
Y^{m,\m}_l~=~(1+x)^{-\frac{(m-\m)}{2}}(1-x)^{\frac{-(m+\m)}{2}}.
\frac{d^{l-m}}{d^{l-m}x}~[(1+x)^{l-\m}(1-x)^{l+\m}]~e^{im\phi}~e^{-i\m\chi}
\eeq
where $x~=~cos{\th}$ and the quantity m and $\m$ just represent the
eigenvalues of the operator $i\frac{\partial}{\partial{\phi}}$ and
$i\frac{\partial}{\partial{\chi}}$ respectively. It is noted that apart from
the usual angles $\th$ and $\phi$, we have an extra angle $\chi$ denoting the
rotational orientation around a specified fixed axis attached to a space-time 
point $x_\m$ giving rise to an anisotropy
in the space-time manifold [12]. The fact that in such an anisotropic space
the angular momentum can take the value 1/2 is found to be analogous to the
result that a monopole charged particle composite representing a dyon
satisfying the condition $e{\m}~=~1/2$ has its angular momentum shifted by
1/2 unit and its statistics shifted accordingly [13]. This suggests that a
fermion can be viewed as a scalar particle moving with $l=1/2$ in an
anisotropic space. The specification of the $l_z$-value for the particle and
antiparticle states then depicts it as a chiral spinor. This may be associated
with a spin system when electrons are polarized in one or the other direction.

 Now we note that when the angle
$\chi$ depicting the rotational orientation around the direction vector
$\xi_{\m}$ attached to the space-time point $x_{\m}$ is gradually changed over
the closed path $0~\le~\chi~\le~2\pi$ it gives rise to the phase factor $\m$ 
in the wave function.  Indeed, the angular part associated with the angle 
$\chi$ in the
spherical harmonics $Y^{m,\m}_l$ is given by $e^{i\m \chi}$ where we have
\beq
i\frac{\partial}{\partial{\chi}} e^{-i\m \chi}~=~\m e^{-i\m \chi}
\eeq
Now when $\chi$ is changed to $\chi+\delta \chi$, we find 
\beq
i\frac{\partial}{\partial{(\chi+\delta \chi)}}e^{-i\m \chi}~=~ 
i\frac{\partial}{\partial{(\chi+\delta \chi)}}
e^{-i\m(\chi+\delta \chi)}~e^{i\m \delta \chi}
\eeq
Thus the wave function will acquire an extra topological phase factor 
$e^{i\m \delta\chi}$
when the angle $\chi$ is changed over the closed path $0~\le~\chi~\le~2\pi$.
For one such complete rotation, the wave function will acquire the phase 
\beq
e^{i\m {\int^{2\pi}_0\delta \chi}}~=~e^{i2\pi \m}
\eeq
Thus for a closed parameter space we have the extra phase factor $e^{i2\pi
\m}$ which represents the Berry phase [14]. The
Berry phase aquired by such a particle is given by 

$$e^{i{\phi}_B}~=~e^{i\m\int^{2\pi}_0{ d\chi}}~=~e^{i2 \pi \m} $$
where
\beq
{\phi}_B~=~2\pi \m
\eeq

    It may be noted  here that the Berry phase is associated with the chiral
anomaly which is caused by quantum mechanical symmetry breaking when a chiral
current interacts with a gauge field [15]. Indeed, the divergence of the axial
vector current in the quantum mechanical case does not vanish and is
associated with the topological quantity known as the Pontryagin index through
the relation 
\beq
q~=~-\frac{1}{16 {\pi}^2}~\int~Tr ^*F_{\m\n}~F_{\m\n}~d^4x
~=~\int~\partial_{\m}J^2_\m d^4x ~=~-\frac{1}{2}~\int \partial_\m~J^5_\m~d^4x
\eeq
where $J^5_\m$ is the axial vector current and $q$ is the Pontryagin index. The
Pontryagin index associated with the integral of the chiral anomaly is related
to the Berry phase which arises when a quantum particle described by a
parameter dependent Hamiltonian moves in a closed path. Indeed, the Berry
phase acquired by such a particle is given by $e^{i{\phi}_B}$ where 
\beq
{\phi}_B~=~2\pi \m~=~\pi q
\eeq
with the relation $q~=~2\m$ [16]. The interction of this axial vector current 
$J^5_\m$ with the non-Abelian  $SL(2,c)$ gauge field $B_\m$, residing on the
lattice link is well known [17].

  It is well known that antiferromagnetism is characterised by strong spin 
fluctuation in 2D which 
causes this strongly correlated electronic system a
parity violated ground state known as $commensurate~ flux~ state$ or $ chiral~
spin~ state$. 
In a recent paper [18] the topological analysis of Heisenberg antiferromagnet 
on a
lattice we have associated the Berry phase with the chirality of the fermions.
The anisotropic antiferromagnetic Heisenberg system can be represented by 
a Hamiltonian

\beq
H~=~J~\sum~(~S^x_iS^x_{j}~+~S^y_iS^y_{j}~+~\Delta ~S^z_iS^z_{j})
\eeq
(with $i$ and $j$ as two-dimensional index)
where $J~>~0$ and $\Delta ~\geq~0$. $\Delta$, the anisotropic parameter is
shown to be
 $\Delta~=~\frac{2\m+1}{2}$ where $\m$ is related
 to the Berry phase
factor.
It may be mentioned here that the relationship of the anisotropic parameter
 $\Delta$
with the Berry phase factor $\m$ has been formulated from an analysis of the
relationship between conformal field theory in 1+1 dimension, Chern-Simon
theory in 2+1 dimension and chiral anomaly in 3+1 dimension. 
 It is noted that for $\m=1/2$ when $\Delta~=~1$ one has the 
antiferromagnetic
Heisenberg model.
Thus in this methodology the antiferromagnetic commensurate flux state 
 will correspond to the Berry phase factor $\m=1/2$ and the corresponding
phase will be given by 
 \beq
{\phi}_B~=~2\pi \m~=~\pi~~~~~~,~~~~~~
\eeq
The parity and time reversal symmetry is spontaneously broken in this
two-dimensional correlated electronic system . 
 This leads to an eqivalent picture of this commensurate flux 
state before doping with the state of the chiral fermions as discussed.

 Let us now consider the state after doping. Due to a hole doping the Berry
phase factor related to antiferromagnetic system will be changed. Doping of a
hole will introduce an extra $\m=1/2$ value to the total topological phase. As
a result the total Berry phase will be given by 
\beq
{\phi}^{\prime}_B~=~2\pi {\m^{\prime}}~=~2\pi (1/2 \pm 1/2)=either~ 0~ or~
 2 \pi.
\eeq
This doped state is a superconducting ground state which can be assured from
topological view point. This state corresponds to the Berry phase factor
$\m^\pr=0~or~1$. 
Regarding the parity and time reversal symmetry we observe that it is restored
in this doped system. We can refer to this change as  a $topological$ $phase$
$transition$ where the Berry phase factor $\m=1/2$ is changed to $\m^\pr=0~or~
1$.
In this connection we should mention here the analysis of Weigmann [19]
where he arrived at the same result in a different way. 

Doping gives rise to superconducting phase coexisting with
a spin glass phase [20] which arises due to frustration of the spin system. In
general the motion of a single hole in an antiferromagnet is frustrated because
it stirs up the spins and creates strings of broken bonds (magnetically
disordered possibly in ferromagnetic phase).
This idea is supported by ARPES experiment [7,8] which found that the bandwidth
of a single hole is controlled by the exchange integral $J$, rather than the
hopping amplitude t [21]. This frustration gives rise to a kinetic driving 
force
for electronic phase separation causing spin-charge separation. In this 
connection the review of Emery and  Kivelson [22] may be mentioned.
This analysis helps us to write the doped antiferromagnetic system as a
composite one  which consists of (i) a 
bosonic system of spinless charges
(holons) and a chargeless spin (spinon) system.

     One thing to be mentioned here that the spin-charge separation state will 
satisfy an important
criteria that the transverse gauge fluctuation will be found suppressed in the
long wavelength and low energy regime so that spinon and holon are deconfined 
at
finite doping (i,e, spin-charge separation). At half filling limit, a long
range AF order can be recovered in 2D and at finite doping, the gauge
fluctuation is shown to be suppressed so that one has a real spin-charge
separation. But there are still residual interactions between spinon and
holon, and spinon can always feel the existence of holon nonlocally (by seeing
the flux quanta bound to holon) through a gauge $A_\m$. So we may consider an 
interaction between real scalar field ${\phi}_n$
(for holon) and chiral spinor field ${\psi}_n$ (for spinon) which are taken to
transform under a rigid (global) gauge transformtion. This interaction
vanishes in 1D which we shall show in a different note and so we may say 
that real spin-charge
separation is visualised.

  On account of the spin charge separation we may
consider that the Berry phase factor $\m~=~0~or~1$ to be concentrated on the
 spinon 
system. There is no contribution of the topological phase factor $\m$ to the
 holons in this superconducting system. 
It is to be noted that after doping superconducting system acquires Berry 
phase due to unbalanced
spin precession in the antiferromagnetic system.

 Now from the angular momentum relation of 2D lattice on 3D sphere of an
anisotropic space 
$$\vec{J}~=~\vec{r}~\times~\vec{p}~-~\m\vec{r}$$ for $|\m|~=~0~or~1$ (or any
integer) we can use a transformation which effectively suggests that we can
have a dynamical relation of the form 
\beq
\vec{J}~=~\vec{r}~\times~\vec{p}~-~\m \vec{r}~~=~~\vec{r^\pr}~\times~
\vec{p^\pr}
\eeq
This equation indicates that the Berry phase which is associated with the
integral value of $\m$ may
be unitarily removed to the dynamical phase. However, to
observe the effect of the Berry phase, we can split the state into a pair of
spinons, each with the constraint of representing the state $\m~=~\pm
1/2$. Now $\m~=~0~or~1$ is achieved for a pair of spinons each having
$\m~=~\pm 1/2$.  As the factor $\m$ is associated with chiral anomaly we may
associate a spinon with the axial vector current $$J^5_{\m}~=~\bar {\psi} 
{\gamma}_{\m} {\gamma}_5 \psi$$. So in our framework we may represent the 
spinon spinon 
interaction by the interction of the
axial vector current $J^5_{\m}$ in eqn (7) through a gauge field $B_{\m}$
which will reside on the lattice link. This justifies the spin pairing of the 
superconducting state. The spin pairs are in a coherent 
phase state as they correspond to the same phase and high
temperature superconductivity is established by the motion of these
coherent pairs which accounts for the high transition temperature of the
cuprates.

  Previously [23] we have shown that the FQH state with $\n=1/2$  
corresponds to the Berry phase factor $\m=1$. So topologically we can
propose an equivalent picture of superconducting state and $\n=1/2$ FQH
state. In a recent publication [24] we have analysed the equivalence 
of RVB states with $\n=1/2$ FQH states in terms of the Berry phase and 
showed that the topological mechanism of 
superconductivity  is analogous to the topological aspects of fractional 
quantum Hall effect with
 $\n=1/2$. This  state is characterised by a flux
$\phi_0~=~\frac{hc}{2e}$ which enables  our theory to satisfy  with one 
experimental
constraint- the order parameter with charge 2e. This effective charge does not
arise from the pairing of nearest neighbour holons. We believe
that these holons are
transported by scattering which arises due to the interacting gauge between 
spinon and holon. Thus the high $T_c$ superconductivity can be naturally
viewed  by inplane kinetic mechanism. Finally we may mention here that this
analysis will be valid for a 3D antiferromagnetic lattice system too and as
such this topological superconductivity is just not restricted to a 2D system
but is also feasible in 3D system also. Detailed investigation along this line
will be pursued later on. 

{\bf Acknowledgement} : Authors are highly grateful to Prof. P.Bandyopadhyay
for helpful discussions. One of the authors (D.P.) is also grateful to the 
Council
of Scientific and Industrial Research (CSIR) for a financial grant 
supporting this research.

\begin{center}
{\bf References}
\end{center}
\begin{enumerate}
\item C.E. Gough et al., Nature, ${\bf 326}$, (1987), 855.
\item N.P. Ong, in $ Physical$ $ Properties$ $ of$ $High$ $ Temperature$
$Superconductors$
edited by P.M. Ginzberg, (World Scientific 1992), vol.3, 159.
\item A.B. Kaiser and C. Uher in $Studies$ $of$ $ High$ $ Temperature$
$Superconductors$, edited by A. Narlikar, (Nora Science Publishers, New York
1991). 
\item C.H. Pennington and C.P. Slichter, in $ Physical$ $ Properties$ $ of$ 
$ High$ 
 $ Temperature$ $ Superconductors$ edited by P.M. Ginzberg, (World Scientific
 1992), vol.2, 269.
\item R.J. Birgeneau G.Shirane, in {\it Physical Properties of 
 High }{\it Temperature Superconductors} edited by 
P.M. Ginzberg, (World Scientific
 1992), vol.1, 151.
\item V.J.Emery and S.A.Kivelson, Phys.Rev.Lett., ${\bf 74}$, (1995), 3253.
\item  J.C.Campuzano et al., Phys.Rev., ${\bf B53}$, (1996), R14737.
\item (a) C.G.Olson et al., Science, ${\bf 245}$, (1989), 731; Physica 
${\bf C162-164 }$, (1989), 1697; Phys.Rev.Lett., ${\bf B42}$, (1990), 381;
\item[](b) Z.-X.Shen et al.,Phys.Rev.Lett.,${\bf 70}$, (1993), 1553; 
\item[](c)H.Ding et al.,Phys.Rev. ${\bf B 54}$,(1996), R9678.
\item B. Basu and P. Bandyopadhyay, Int. J. Mod. Phys. ${\bf B, 12}$ (1998), 
2649.
\item M. Fierz, Helv. Phys. Acta. ${\bf 17}$ (1944), 27.
\item C.A. Hurst, Ann. Phys. ${\bf 50}$ (1968), 51.
\item (a)B. Basu, Mod. Phys. Lett. ${\bf B, 6}$ (1992), 1601.
\item[] (b)P. Bandyopadhyay, Int. J. Mod. Phys. ${\bf A4}$ (1989), 4449.
\item F.D.M. Haldane, Phys. Rev. Lett. ${\bf 51}$ (1983), 605.
\item  D. Banerjee and P. Bandyopadhyay, J. Math. Phys. ${\bf 33}$ (1992), 990.
\item B. Basu and P. Bandyopadhyay, Int. J. Mod. Phys. ${\bf B, 11}$ (1997), 
2707.
\item D. Banerjee and P. Bandyopadhyay, Mod. Phys. Lett. ${\bf B, 8}$ (1994),
 1643.
\item A.Roy and P. Bandyopadhyay, J. Math. Phys. ${\bf 30}$, (1989), 2366.
\item P. Bandyopadhyay, Conformal Field Theory, Quantum Group and Berry Phase
 Int. J. Mod. Phys. {\bf A}(accepted for publication).
\item P.B.Wiegmann, Phys. Rev. Lett. ${\bf 65}$, (1990), 2070. 
\item (a)A.Weidinger et al., Phys. Rev. Lett. ${\bf 62}$, (1989), 102.
\item[] (b) Ch. Niedermayer et al.,  Phys. Rev. Lett. ${\bf 80}$, (1998), 3843.
\item B.O. Wells et al., Science ${\bf 277}$, (1997), 1067.
\item S.A. Kivelson and V.J. Emery in $Strongly$ $Correlated$ $Electronic$
$Materials:$ $The$ $ Los$ $ Alamos$ $Symposium$ $1993$
 edited by K.S.Bedell et al,
(Addision-Wesley, Reading Massachusetts, 1994), 619.
\item B. Basu, D. Banerjee and P. Bandyopadhyay, Phys. Lett. ${\bf A, 236}$
 (1997), 125.
\item B. Basu, D.Pal and P. Bandyopadhyay, Int. J. Mod. Phys. ${\bf B, 13}$
(1999), 3393.

\end{enumerate}

\end{document}